\documentclass[aps,pra,twocolumn]{revtex4-2}
\usepackage{color}
\usepackage{longtable}
\usepackage{dcolumn}
\usepackage{graphicx}
\usepackage{bm}
\usepackage{bbm}
\usepackage{times}
\usepackage{nicefrac}
\usepackage{amsmath}
\usepackage{amsfonts}
\usepackage{amssymb}
\usepackage{amsthm}

\newcommand{\lbr}{\langle}
\newcommand{\rbr}{\rangle}

\allowdisplaybreaks

\makeatletter
\DeclareRobustCommand{\cev}[1]{%
  \mathpalette\do@cev{#1}%
}
\newcommand{\do@cev}[2]{%
  \fix@cev{#1}{+}%
  \reflectbox{$\m@th#1\vec{\reflectbox{$\fix@cev{#1}{-}\m@th#1#2\fix@cev{#1}{+}$}}$}%
  \fix@cev{#1}{-}%
}
\newcommand{\fix@cev}[2]{%
  \ifx#1\displaystyle
    \mkern#23mu
  \else
    \ifx#1\textstyle
      \mkern#23mu
    \else
      \ifx#1\scriptstyle
        \mkern#22mu
      \else
        \mkern#22mu
      \fi
    \fi
  \fi
}
\makeatother

\begin{document}
\preprint{Version 1.0}

\title{Finite nuclear mass  correction to the hyperfine splitting in hydrogenic systems}

\author{Krzysztof Pachucki}

\affiliation{Faculty of Physics, University of Warsaw,
             Pasteura 5, 02-093 Warsaw, Poland}

\date{\today}
\begin{abstract}
A general quantum electrodynamic method for the derivation of nuclear recoil corrections in hydrogenic systems, which are exact in the nuclear charge parameter $Z\,\alpha$, is introduced.
The exemplary derivation is presented for the $O(m/M)$ nuclear pure recoil correction  to the hyperfine splitting.
The obtained result is verified by comparison to the known $(Z\,\alpha)^5$ contribution.
\end{abstract}


\maketitle

\section{Introduction}

Consider a two-body system with an arbitrary nucleus and a point light fermion, such as an electron or muon.
If the nucleus can be treated as a static source of an electric potential, then
its energy levels can be  obtained from the Dirac equation
\begin{align}
H_D\,\phi = E_D\,\phi\,, \label{01}
\end{align}
where
\begin{align}
H_D = \vec\alpha\cdot\vec p + \beta m +V_C \label{02}
\end{align}
and where $V_C$ is a Coulomb potential  including the nuclear charge distribution $\rho_C(r)$,
\begin{align}
V_C(r) =&\ - \int d^3r'\,\frac{Z\,\alpha}{|\vec r-\vec r\,'|}\,\rho_C(r')\,. \label{03}
\end{align}
The Dirac equation is valid only in the limit of the infinite nuclear mass $M$, and
there is no corresponding equation for the finite nuclear mass case.
This means that we are not able to treat exactly two-body systems with arbitrary masses
in the relativistic theory, in contrast to nonrelativistic quantum mechanics.
 
 There are in principle two perturbative approaches that are employed for two-body systems.
 The first relies on expansion in powers of $Z\,\alpha$ while keeping an arbitrary mass ratio
 \begin{align}
 E\Big(\frac{m}{M}, Z\,\alpha\Big) =&\ m+M + E^{(2)} + E^{(4)} + E^{(5)} + E^{(6)} + \ldots \label{04}
 \end{align}
 where $E^{(n)}$ is of the order $(Z\,\alpha)^n$ and may sometimes contain finite powers of $\ln(Z\,\alpha)$.
In this expansion, the coefficient  $E^{(2)}$ is the nonrelativistic energy,  
 $E^{(4)}$ is the relativistic correction, and $E^{(n>4)}$ are higher-order quantum electrodynamics (QED) and relativistic  corrections. 
 These corrections for point arbitrary mass particles have already been obtained up to $E^{(6)}$ \cite{adkins_2023, twobody, ptwobody}
 using the so-called nonrelativistic QED (NRQED) approach.
For the finite nuclear size case, they are also known up to  $E^{(6)}$  for an arbitrary mass ratio \cite{twobody, ptwobody},
with the exception of $S$-states  which are known only in the nonrecoil limit \cite{threephoton}.
 
 In the second approach, one performs an expansion in the mass ratio $m/M$,
 while keeping the parameter $Z\,\alpha$ arbitrary,
 \begin{align}
 E\Big(\frac{m}{M},Z\,\alpha\Big) = E_D(Z\,\alpha) + E_\mathrm{rec}(Z\,\alpha) + \ldots\,, \label{05}
 \end{align}
 where the Dirac energy $E_D$ is in the infinite nuclear mass limit, $E_\mathrm{rec}$
 is the first order in the mass ratio $\sim m/M$ correction, while the higher-order terms are merely unknown.
 This first order in mass ratio correction (for the point nucleus) was first derived  Shabaev in Refs. \cite{shabaev:85,shabaev:88}.
Next, it was independently rederived  in Ref. \cite{pachucki:95}, 
 together with the calculations of the, unknown at that time, $(Z\,\alpha)^6\,m^2/M$ correction.
 The compact form for this first-order recoil correction for a point nucleus was obtained in Ref. \cite{yelkhovsky:96}. 
 Soon after, the direct numerical calculations were performed in Refs.~\cite{artemyev:95:pra,artemyev:95:jpb}.
 The generalization for the finite size nucleus was achieved only recently in Ref. \cite{fsrec}, and the formulas are
 the following,
\begin{align}
E_{\rm rec} = \frac{i}{M} \int_{-\infty}^{\infty}
  &\
\frac{d\omega}{2\,\pi}\,
 \lbr \phi | \big[ p^j - D_C^j(\omega)\big] \,
  \nonumber \\ & \times
 G(E_D + \omega) \,
 \big[ p^j - D_C^j(\omega)\big] | \phi \rbr\,, \label{06}
\end{align}
where  $\phi$ is an eigenstate of the Dirac-Coulomb Hamiltonian in Eq. (\ref{02}),
\begin{align}
G(E) = [E-H_D(1-i\epsilon)]^{-1}
\end{align}
is the Dirac-Coulomb Green's function, $D_C^j(\omega)$ is an operator which in coordinate representation
is given by the function
\begin{align}
D_C^j(\omega,\vec r) =&\ -4\pi Z\alpha \, \alpha^i \, G_{C}^{ij}(\omega,\vec{r})\,,
\end{align}
where
\begin{align}
G_C^{ij}(\omega,\vec r) =&\ \int \frac{d^3k}{(2\,\pi)^3}\,\frac{1}{k^2}\,\biggl(\delta^{ij}-\frac{k^i\,k^j}{\vec k\,^2}\biggr)\,e^{i\,\vec k\,\vec r}
\end{align}
is the transverse photon propagator for a point nucleus, and 
\begin{align}
G_C^{ij}(\omega,\vec r) =&\ \int \frac{d^3k}{(2\,\pi)^3}\,\biggl[\frac{\rho_C(-k^2)}{k^2}\,\biggl(\delta^{ij}-\frac{k^i\,k^j}{\omega^2}\biggr) 
\nonumber \\ &\
- \frac{k^i\,k^j}{\omega^2}\,\frac{\rho_C(\vec k^2)}{\vec k^2}\biggr] \,e^{i\,\vec k\,\vec r} \label{10}
\end{align}
for the finite-size nucleus \cite{fsrec}, with $ k^2 = \omega^2-\vec k\,^2 $ and $\omega = k^0$.
Equation (\ref{06}) will be further transformed to two different forms. For convenience, we will give different names
for different forms of this recoil correction to energy, and this one in Eq. (\ref{06}) we call the ${\cal F}_1$ form.

In this papers we derive the analogous formula for the leading recoil correction  to the hyperfine splitting (hfs) $E_\mathrm{hfsrec}$
and verify it by calculation of the $(Z\,\alpha)^5$ contribution. This formula has a very universal character,
such as that in Eq. (\ref{06}), and it can be used for an analytic derivation of the $Z\,\alpha$ expansion or for direct numerical calculation.  
In view of the planned hyperfine splitting measurements in muonic atoms \cite{Aldo}, this nonperturbative numerical
calculation of $E_\mathrm{hfsrec}$ would be helpful  in the interpretation of  the measured hfs in terms of the nuclear magnetization distribution.
The accurate recoil corrections  to hfs are important also in view of significant disagreements for $\mu$D \cite{muD} and $^6$Li \cite{7Li_theory, Patkoshfs, Drakehfs} measured values.

\section{Recoil correction in the temporal gauge}
We first transform Eq. (\ref{06}) to a different ${\cal F}_2$ form 
and introduce a notation $\int_s$ for a symmetric integration around a pole at $\omega=0$,
\begin{align}
E_{\rm rec} =&\ -\frac{i}{M} \int_s \frac{d\omega}{2\,\pi}\, \lbr \phi | \big[ p^j(V_C) - \omega\,D_C^j(\omega)\big] 
\nonumber \\ &\times
G(E_D + \omega ) \, \big[ p^j(V_C) + \omega\,D_C^j(\omega)\big] | \phi \rbr\, \frac{1}{\omega^2}\,,
\end{align}
where $p^j(V_C) = [p^j\,,\,V_C]$.
Let us introduce a photon propagator in the temporal gauge including the finite nuclear size \cite{ulj},
\begin{align}
G_T^{ij}(\omega, \vec k) =&\ \frac{\rho_C(-k^2)}{k^2}\,\biggl(\delta^{ij}-\frac{k^i\,k^j}{\omega^2}\biggr)\,, \label{12}
\end{align}
and analogously
\begin{align}
D_T^j(\omega,\vec r) =&\ -4\pi Z\alpha \, \alpha^i \, G_{T}^{ij}(\omega,\vec{r})\,. \label{13}
\end{align}
The temporal  gauge is a particular case of the axial gauge and is defined by the condition
$G^{0\mu}_T = 0$. The relation to the propagator in the Coulomb gauge is 
\begin{align}
D_C^j(\omega) =&\  D_T^j(\omega)  + \frac{1}{\omega^2}\,\big[\omega+E_D-H_D\,,\,p^j(V_C)\big].
\end{align}
Using $D_T$, the recoil correction $E_\mathrm{rec}$ takes the form 
\begin{align}
E_\mathrm{rec} =&\ \frac{i}{M} \int_s \frac{d\omega}{2\,\pi}\, \lbr \phi | D_T^j(\omega) \,G(E_D + \omega)\,D_T^j(\omega) | \phi \rbr\,. \label{14}
\end{align}
The simplest form, called here the ${\cal F}_3$ form,  for the recoil correction is achieved when using the temporal gauge for the photon propagator.
We will use this observation when presenting the recoil correction to the hyperfine splitting.

\section{Derivation of recoil correction to the binding energy}
The original derivation \cite{shabaev:85,  shabaev:88, pachucki:95} of the nonperturbative formula for the recoil correction was quite complicated.
Here, we present a very much simplified derivation, which later will be used for the hyperfine splitting.

Consider the nonrelativistic kinetic energy of the nucleus 
\begin{align}
H_N = \frac{1}{2\,M}\,\big[\vec P-q\, \vec A(\vec R)\big]^2\,,
\end{align}
where $\vec P = -i\vec\nabla_R$, $q = -Z\,e$, and $e$ is the electron charge. 
The leading recoil correction can formally be written as the expectation value
\begin{align}
E_\mathrm{rec} =&\ \frac{1}{2\,M}\,\langle\Psi| (\vec P-q\,\vec A)^2|\Psi\rangle_\mathrm{QED} 
\end{align}
on a hydrogenic state $|\Psi\rangle_\mathrm{QED}$ (which is centered at the position of nucleus $\vec R$) in quantum electrodynamic  (QED) theory. 
The meaning of this expectation value is not obvious and is explained as follows.
The matrix element of an arbitrary operator $Q$ on a state $\Psi$ is
\begin{align}
\langle\Psi|Q|\Psi\rangle_\mathrm{QED} = 
\frac{\langle\Psi|{\mathrm T} Q\,\exp[-i\int d^4y\, H_I(y)]|\Psi\rangle}{\langle\Psi|{\mathrm T}\exp[-i\int d^4y\, H_I(y)]|\Psi\rangle}\,, \label{17}
\end{align}
where T denotes chronological ordering with an assumption that the time coordinate of $Q$ is $t=0$,
the interaction Hamiltonian is 
\begin{align}
H_I(y) = e\,j_\mu(y) A^\mu(y)\,, 
\end{align}
and $|\Psi\rangle$ is the bare hydrogenic state in the second quantized theory.
We keep in mind  that in Eq. (\ref{17}), for the purpose of this work, 
all the electron self-energy and vacuum polarizations are being neglected.
The crucial point is the interpretation of $\vec P$ and its action on $|\Psi\rangle_\mathrm{QED}$. 
Namely, consider the representation of the fermion field
in terms of creation and annihilation operators of one-particle hydrogenic states $\phi_s$,
\begin{align}
\hat\psi(x)=&\ \sum_s^+ a_s\phi_s(\vec x)\,e^{-i\,E_s t} + \sum_s^- b_s\phi_s(\vec x)\,e^{-i\,E_s t}\,,
\nonumber \\
\hat\psi^+(x)=&\ \sum_s^+ a^+_s\phi^+_s(\vec x)\,e^{i\,E_s t} + \sum_s^- b^+_s\phi^+_s(\vec x)\,e^{i\,E_s t}\,;
\end{align}
for details, see Appendix B. The differentiation $\vec\nabla_R$  acts on functions $\phi_s$ and 
operators $a_s, b_s$, and this can be represented as
\begin{align}
\vec \nabla_R =&\ \int d^3r\, \hat \psi^+(\vec r)\,\vec \partial_R\,\hat \psi(\vec r) + \vec\partial_R
\nonumber \\ =&\
-\int d^3r\, \hat \psi^+(\vec r)\,\vec \partial_r\,\hat \psi(\vec r) + \vec\partial_R \label{20}
\end{align}
where $\hat\psi(\vec r) \equiv \hat\psi(0,\vec r)$, and $\vec\partial_R$ is understood in the following sense.
The hydrogenic state $\phi_s$ is a function of $\phi_s(\vec r -\vec R)$ of the difference in electron  and nucleus position vectors, 
therefore  $\vec \partial_R\,\phi_s = -\vec \partial_r\,\phi_s$, and $\hat a_s, \hat b_s$ remain intact.
As a test, for $t=0$,
\begin{align}
\vec\nabla_R\hat\psi(0, \vec x) =&\ -\int d^3r\, \hat \psi^+(\vec r)\,\vec \partial_r\,\hat \psi(\vec r)\;\hat\psi(0, \vec x) - \vec\partial_x\hat\psi(0, \vec x)  
\nonumber \\ =&\  0\,,
\end{align}
as it should. Moreover, for an arbitrary Fock state $|\Psi\rangle$, 
\begin{align}
\vec \nabla_R\,|\Psi\rangle = -\int d^3r\, \hat \psi^+(\vec r)\,\vec \partial_r\,\hat \psi(\vec r)|\Psi\rangle\,,
\end{align}
and this holds in particular for the vacuum state $|0\rangle$.

We can now return to the expectation value of the nuclear kinetic energy, and we split it into three parts,
\begin{align}
E_\mathrm{rec} =&\ \frac{1}{2\,M}\,\langle\Psi| (\cev \nabla_R+i\,q\,\vec A) (\vec \nabla_R-i\,q\,\vec A)|\Psi\rangle_\mathrm{QED} 
\nonumber \\ =&\ E_C+E_T+E_S\,,
\end{align}
where the quadratic derivative is understood in the symmetric form, namely
\begin{align}
\vec\nabla_R^2 = -\cev{\nabla}_R\,\vec\nabla_R\,.
\end{align}
Let us start the derivation from the Coulomb part $E_C$, where $|\Psi\rangle_\mathrm{QED}$ can be replaced by $|\Psi\rangle$ 
because we neglect all the radiative corrections.
The hydrogenic state $|\Psi\rangle = \hat a^+_\phi|0\rangle$, and
\begin{align}
E_C =&\ \frac{1}{2\,M}\langle \Psi|\vec P^{\,2}|\Psi\rangle 
\nonumber \\ =&\  
\frac{1}{2\,M}\int d^3x'\,d^3x\, \langle 0|\hat a_\phi\, (\hat \psi^+\,\vec p\,\hat \psi)_{x'}\,(\hat \psi^+\,\vec p\,\hat \psi)_x\,\hat a^+_\phi| 0\rangle
\nonumber \\ =&\  
\frac{1}{2\,M} \int d^3x'\,d^3x\,\phi^*(x')\,\vec p\,'\, \bigl[\langle 0| \hat \psi(x')\,\hat \psi^+(x)|0\rangle
\nonumber \\ &\
 - \langle 0| \hat \psi^+(x)\,\hat \psi(x')|0\rangle\bigr]\, \vec p\,\phi(x)
\nonumber \\ =&\  
\frac{1}{M} \int d^3x'\,d^3x\,\phi^*(x')\,\vec p\,'\,{\mathrm T}\langle 0| \hat \psi(x')\,\hat \psi^+(x)|0\rangle\, \vec p\,\phi(x)
\nonumber \\ =&\  \frac{1}{2\,M} \langle\phi|\,\vec p\,(P_+-P_-)\,\vec p\,|\phi\rangle
\nonumber \\ =&\  \frac{i}{M} \int\frac{d\omega}{2\,\pi}\,\langle\phi|\,p^j\,G(\omega + E_0)\,p^j\,|\phi\rangle\,.
\end{align}
We note that the leading recoil correction is obtained from the above by replacing $P_+=I-P_-$
and subsequent neglect of $P_-$, thus 
\begin{align}
E_\mathrm{rec} \approx \langle\phi| \frac{\vec p\,^2}{2\,M}|\phi\rangle. \label{27}
\end{align}

The second  part is the single transverse photon exchange,
\begin{align}
E_{T} =&\ \frac{1}{2\,M}\,\langle\Psi| \{\vec P\,,\,Z\,e\,\vec A(\vec R)\}|\Psi\rangle_\mathrm{QED} 
\nonumber \\ =&\ 
-\frac{i\,Z\,e}{2\,M}\,\langle\Psi| A^i(\vec R)\,\nabla^i_R|\Psi\rangle_\mathrm{QED} +\mathrm{h.c.}
\nonumber \\ =&\ 
\frac{Z\,e^2}{2\,M}\,\int d^4 y\,\langle\Psi| {\mathrm T}\big[A^i(\vec R)\,\nabla^i_R\,j^j(y)\,A^j(y)\big] |\Psi\rangle +\mathrm{h.c.}
\nonumber \\ =&\ 
\frac{i\,Z\,e^2}{2\,M}\,\int d^4y\,\langle\Psi| {\mathrm T}\big[\nabla^i_R\,j^j(y)\big] |\Psi\rangle\, G_C^{ij}(y-R) +\mathrm{h.c.}
\nonumber \\ =&\ E_{T1}  + E_{T2} \,,
\end{align}
where the $\vec\nabla_R$ operator is assumed at $t=0$ in the chronological ordering. 
$E_{T1}$ is due to the first term in Eq. (\ref{20}), so
\begin{align}
E_{T1} =&\  
-\frac{i\,Z\,e^2}{2\,M}\,\int d^4y \, G_C^{ij}(y-R)\int d^3x\, 
\nonumber \\ &\times 
\langle 0|\hat a_\phi\, {\mathrm T}[(\hat \psi^+\partial^i\,\hat \psi)_x \,(\hat \psi^+\alpha^j\,\psi)_y]\,\hat a^+_\phi| 0\rangle +\mathrm{h.c.}
\nonumber \\ =&\
-\frac{i}{M}\, \int\frac{d\,\omega}{2\,\pi} \Bigl[\langle\phi|p^j\,G(\omega+E_0)\,D_C^j(\omega)|\phi\rangle 
\nonumber \\ &\ 
+ \langle\phi| D_C^j(\omega)\,G(\omega+E_0)\, p^j |\phi\rangle \Big] \,.
 \end{align}
 $E_{T2}$, due to the second term in Eq. (\ref{20})
 \begin{align}
E_{T2} =&\  \frac{i\,Z\,e^2}{2\,M}\,\int d^4y\,\Theta(-y_0)\,\langle\Psi| j^j(y) |\Psi\rangle\, 
\nonumber \\ &\times
\partial^i_y G^{ij}(y-R) +\mathrm{h.c.} = 0\,,
\end{align}
vanishes due to the current conservation $\langle\Psi| \partial^i_y\,j^i(y) |\Psi\rangle = 0$.

The third part, the double transverse (seagull) contribution $E_S$,  is
\begin{align}
E_S =&\ \frac{Z^2\,e^2}{2\,M}\,\langle \Psi| \vec A^{\,2}(\vec R) |\Psi\rangle_\mathrm{QED}
 \nonumber \\ =&\ 
 \frac{Z^2\,e^2}{2\,M}\,\frac{(i\,e)^2}{2}\,\int d^4x \int d^4 y
 \nonumber \\ &\times
\langle \Psi| {\mathrm T}\big[\vec A^{\,2}(\vec R)\,\vec j(x)\cdot \vec A(x)\; \vec j(y)\cdot\vec A(y)\big] |\Psi\rangle
\nonumber \\ =&\ 
\frac{i}{M}\,\int\frac{d\omega}{2\,\pi}\,\langle\phi| D_C^j(\omega)\,G(\omega+E_0)\,D_C^j(\omega)\,\alpha^j|\phi\rangle\,.
\end{align}
The sum $E_C + E_T + E_S$ gives $E_\mathrm{rec}$  in the ${\cal F}_1$ form in Eq. (\ref{06}), 
which is next transformed to the ${\cal F}_3$ form in Eq. (\ref{14}).
The same approach will be used for the derivation of the recoil correction to the hyperfine splitting.

\section{Nonperturbative recoil correction to HFS}
In the relativistic formalism the hyperfine splitting for the point and infinitely heavy nucleus is obtained from the expectation value of
\begin{align}
V_{\rm hfs} =  -e\,\vec\alpha\cdot\vec A_I \,, 
\end{align}
where
\begin{align}
e\,\vec A_I(\vec r) = \frac{e}{4\,\pi}\,\vec\mu\times\frac{\vec r}{r^3}\,,
\end{align}
on a state $\phi$
\begin{align}
E_{\rm hfs} =  \langle \phi| V_{\rm hfs} |\phi\rangle\,.
\end{align}

For the finite size nucleus, the Coulomb interaction becomes
\begin{align}
\biggl[\frac{1}{r}\biggr]_\mathrm{fs} = \int\frac{d^3q}{(2\,\pi)^3}\,4\,\pi\,\frac{\rho_C(\vec q^{\,2})}{\vec q^{\,2}}\,e^{i\,\vec q\,\vec r}
\end{align} 
[cf. Eq. (\ref{03})], and the magnetic one becomes
\begin{align}
\biggl[\frac{\vec r}{r^3}\biggr]_\mathrm{fs} = -\vec\nabla \int\frac{d^3q}{(2\,\pi)^3}\,4\,\pi\,\frac{\rho_M(\vec q^{\,2})}{\vec q^{\,2}}\,e^{i\,\vec q\,\vec r}\,.
\end{align} 
In the following we will assume $\rho_C=\rho_M = \rho$ to keep the notation short, and the final formulas
will later be generalized to $\rho_C\neq\rho_M$.

The recoil correction to the hyperfine splitting  is obtained using the following effective Hamiltonian for a particle 
with an arbitrary spin $I$ and charge $q$, which includes all spin-dependent terms up to $1/M^2$,
\begin{align}
H_\mathrm{nuc} =&\ \frac{\vec\Pi^2}{2\,M} + q\,A^0 -\frac{q}{2\,M}\,g\,\vec I\cdot\vec B 
\nonumber \\ &\
-\frac{q}{4\,M^2}\,(g-1)\,\vec I\cdot[\vec E\times\vec\Pi-\vec\Pi\times\vec E]\,, \label{38}
\end{align}
where $\vec\Pi = \vec P-q\,\vec A$, and where we introduced the nuclear $g$ factor,
\begin{align}
\vec\mu = \frac{q}{2\,M}\,g\,\vec I
\end{align}
We will use this Hamiltonian for the nucleus, where we assume that the charge of the nucleus is $q = -Z\,e$ with $e$ being the electron charge.
Electromagnetic  form factors are neglected in the above,  because they depend on $q_0$ through $q^2=q_0^2-\vec q^{\,2}$ and thus
cannot be included on the Hamiltonian level. 
Because every photon exchange between the nucleus and the electron involves the photon propagator multiplied by the 
nuclear form factors, we can move these form factors to the redefined photon propagators [see Eqs. (\ref{10}) and  (\ref{12})].
The neglect of $q_0$ in nuclear form factors is a common mistakes in relativistic atomic structure calculations.

Using Eq. (\ref{38}) the recoil correction to hfs  is split into three parts,
 \begin{align}
 E_\mathrm{hfsrec} = E_\mathrm{kin} +  E_\mathrm{so} + E_\mathrm{sec}\,,
 \end{align}
which are calculated one by one in the following. 

\subsection{Kinetic energy contribution}
The kinetic energy contribution is
\begin{align}
E_\mathrm{kin} = \frac{1}{2\,M}\,\langle \Psi| \vec\Pi^2  |\Psi\rangle_\mathrm{QED}\,.
\end{align}
For its derivation we use Eq. (\ref{17}) with $H_I$ including the nucleus magnetic interaction
 \begin{align}
H_I(y) = e\,j_\mu(y) A^\mu(y) -\vec\mu\cdot\vec B(y)\,\delta^3(\vec y-\vec R)\,,
\end{align}
and split it into two parts
\begin{align}
E_\mathrm{kin}  = E_\mathrm{kin1}  + E_\mathrm{kin2} \,.
\end{align}
$E_{\rm kin1}$  due to the first term in Eq. (\ref{20}) is obtained from the previous result for the recoil correction to energy,
 \begin{align}
E_{\rm kin1} =&\ E_{C1}+E_{T1}+E_{S1}
  \nonumber \\ =&\ 
\delta_\mathrm{hfs}\,\frac{i}{M} \int \frac{d\omega}{2\,\pi}\,  \lbr \phi | \big[ p^j - D_C^j(\omega)\big]  
  \nonumber \\ & \times
 G(E_D + \omega) \,\big[ p^j - D_C^j(\omega)\big] | \phi \rbr\,,
\end{align}
where the state $\phi$  and the propagator $G$ are corrected by the hyperfine interaction $V_\mathrm{hfs}$ including the finite nuclear size,
namely
\begin{align}
E_{\rm kin1} =&\ \frac{i}{M} \int \frac{d\omega}{2\,\pi} \Bigl[ \lbr \phi | \big[ p^j - D_C^j(\omega)\big]  
 G(E_D + \omega)
  \nonumber \\ &\times 
 \big(V_\mathrm{hfs}  - \langle V_\mathrm{hfs}\rangle\big)
  G(E_D+\omega)\, 
 \big[ p^j - D_C^j(\omega)\big] | \phi \rbr
 \nonumber \\  &\
 +2\, \lbr \phi | V_\mathrm{hfs}\,G'(E_D)\, \big[ p^j - D_C^j(\omega)\big]  
 G(E_D+\omega)
  \nonumber \\ &\times
 \big[ p^j - D_C^j(\omega)\big] | \phi \rbr\Bigr]\,. \label{41}
 \end{align}
$E_{\rm kin2}$ is due to the second term  in Eq. (\ref{20}) and is split into three parts,
\begin{align}
E_\mathrm{kin2} = E_{C3} + E_{C2} + E_{T2}\,, \label{42}
\end{align}
which are calculated as follows. The first part is 
\begin{align}
E_{C3} =&\ \frac{1}{2\,M}\,\langle\Psi| \cev{\partial}_R\,\vec{\partial}_R|\Psi\rangle_\mathrm{QED}
\nonumber \\ =&\ 
\frac{1}{2\,M}\! \int\!  d^4y \! \int \! dt \, \langle\Psi| \mathrm{T}\big[(i\,e\,\vec{A}\,\vec j)_y\,(i\,\vec\mu\vec B)_{t,R}\;\cev{\partial}_R\,\vec{\partial}_R\big] |\Psi\rangle
\nonumber \\ =&\ 
-\frac{i}{M}\,\int_s \frac{d\omega}{2\,\pi} \langle\phi| \nabla^2\,V_\mathrm{hfs}(\omega) |\phi\rangle\,\frac{1}{\omega^2}\,,
\end{align}
where we introduced the frequency-dependent hyperfine interaction
\begin{align}
V_\mathrm{hfs}(\omega,\vec r) =&\   \epsilon^{ijl}\, e\,\mu^i\, \alpha^j\, \partial^l D(\omega, r)\,, \label{46}
\end{align}
such that $V_\mathrm{hfs}(0,r) = V_\mathrm{hfs}(r)$, and
\begin{align}
{D}(\omega,r) = \int \frac{d^3k}{(2\pi)^3}\, e^{i\vec{k}\cdot\vec{r}}\,\frac{\rho({\vec k}^2-\omega^2)}{\omega^2-{\vec k}^2}\,. \label{47}
\end{align}
The second part in Eq. (\ref{42}) is
\begin{widetext}
\begin{align}
E_{C2} =&\ \frac{1}{2\,M}\,\int d^3r\,\langle\Psi| \hat\psi^+(\vec r)\,\vec\partial_r\hat\psi(\vec r)\,\vec{\partial}_R|\Psi\rangle_\mathrm{QED} + \mathrm{h.c.}
\nonumber \\ =&\ 
\frac{1}{2\,M}\,\int d^4y\int dt\,\int d^3r\, \langle\Psi| \mathrm{T}\big[(i\,e\,\vec{A}\,\vec j)_y\,(i\,\vec\mu\,\vec B)_{t,R}\,
\big(\hat\psi^+ \vec\partial\, \hat\psi\big)_r\,\vec{\partial}_R\big]|\Psi\rangle + \mathrm{h.c.}
\nonumber \\ =&\ 
-\frac{i}{M}\int \frac{d\,\omega}{2\,\pi}\,\frac{1}{\omega}\, 
\Bigl[ \langle \phi| \partial^k (V_\mathrm{hfs}(\omega)-V_\mathrm{hfs})\,G(E_0+\omega) |\partial^k\phi\rangle
+ \langle\partial^k\phi| G(E_0+\omega)\,\partial^k (V_\mathrm{hfs}(\omega)-V_\mathrm{hfs}) |\phi\rangle\Bigr]\,.
\end{align}
The third part in Eq. (\ref{42}) is
\begin{align}
E_{T2} =&\ 
-\frac{i\,Z\,e}{2\,M}\,\langle\Psi| A^i(\vec R)\,\partial^i_R|\Psi\rangle_\mathrm{QED} +\mathrm{h.c.}
\nonumber \\ =&\ 
-\frac{iZ\,e}{2\,M}\,\frac{1}{2}\int d^4 y\,d^4 x\,dt\,{\mathrm T}\langle\Psi| A^i(\vec R)\,\partial^i_R\,(i\,e)\,j^j(x)\,A^j(x)\,(i\,e)\,j^k(y)\,A^k(y)\,
(i)\,\mu^l\,B^l(t,\vec R)|\Psi\rangle +\mathrm{h.c.}
\nonumber \\ =&\ 
-\frac{1}{M}\, \int_s \frac{d\omega}{2\,\pi} \,\bigl[
\langle \phi | D_C^i(\omega)\,G(E_0-\omega)\,\partial^i (V_\mathrm{hfs}(\omega)-V_\mathrm{hfs}) |\phi\rangle
+ \langle\phi |\partial^i (V_\mathrm{hfs}(\omega)-V_\mathrm{hfs})\,G(E_0+\omega)\,D_C^i(\omega) |\phi\rangle \bigr]\,\frac{1}{\omega} \,.
\end{align}
Combining these three parts together, $E_\mathrm{kin2}$ in Eq. (\ref{42}) becomes
\begin{align}
E_\mathrm{kin2}=&\
-\frac{i}{M}\,\int_s \frac{d\omega}{2\,\pi} \langle\phi| \partial^j\partial^j (V_\mathrm{hfs}(\omega)-V_\mathrm{hfs}) |\phi\rangle\,\frac{1}{\omega^2}
 - \frac{1}{M}\int_s \frac{d\,\omega}{2\,\pi}\,\frac{1}{\omega}\, 
\Bigl[ -\langle \phi| \partial^j (V_\mathrm{hfs}(\omega)-V_\mathrm{hfs})\,G(E_0+\omega)
\nonumber \\ &\times
\big(p^j-D_C^j(\omega)\big) |\phi\rangle
+\langle \phi| \big(p^j-D_C^j(\omega)\big)\,G(E_0+\omega)\,\partial^j (V_\mathrm{hfs}(\omega)-V_\mathrm{hfs}) |\phi\rangle \Bigr]\,,
\end{align}
and this form we will call ${\cal F}_1$ in analogy to the previous case. It can be further transformed to the ${\cal F}_2$ form,
\begin{align}
E_\mathrm{kin2} =&\
-\frac{1}{M}\int\frac{d\,\omega}{2\,\pi}\,\frac{1}{\omega^2}\, 
\Bigl\{ \langle \phi| \partial^j(V_\mathrm{hfs}(\omega)-V_\mathrm{hfs})\,G(E_0+\omega)\,\big[p^j(V) + \omega\,D_C^j(\omega)\big] |\phi\rangle
\nonumber \\ &\
+\langle \phi| \bigl[p^j(V)-\omega\,D_C^j(\omega)\bigr]\,G(E_0+\omega)\,\partial^j(V_\mathrm{hfs}(\omega)-V_\mathrm{hfs}) |\phi\rangle \Bigr\}
\end{align}
and combined with $E_\mathrm{kin1}$ in the ${\cal F}_2$ form, 
\begin{align}
E_\mathrm{kin1} =&\
-\delta_\mathrm{hfs} \frac{i}{M} \int_{-\infty}^{\infty}\frac{d\omega}{2\,\pi}\, \frac{1}{\omega^2}\,
\lbr \phi | \big[ p^j(V) - \omega\,D_C^j(\omega)\big] \,G(\omega + E_0) \, \big[ p^j(V) + \omega\,D_C^j(\omega)\big] | \phi \rbr
\end{align}
to obtain a simpler expression for $E_\mathrm{kin}$:
\begin{align}
E_\mathrm{kin} =&\
-\delta_\mathrm{hfs} \frac{i}{M} \int_s \frac{d\omega}{2\,\pi}\, \frac{1}{\omega^2}\,
\lbr \phi | \big[ p^j(V_C+V_\mathrm{hfs}(\omega)) - \omega\,D_C^j(\omega)\big] 
\,G(\omega + E_0) \, \big[ p^j(V_C+V_\mathrm{hfs}(\omega)) + \omega\,D_C^j(\omega)\big] | \phi \rbr\,,
\end{align}
which can be further simplified in the ${\cal F}_3$ form
\begin{align}
E_\mathrm{kin} =&\
-\delta_\mathrm{hfs} \frac{i}{M} \int_s \frac{d\omega}{2\,\pi}\, \frac{1}{\omega^2}\,
\lbr \phi | \big[ p^j(V_\mathrm{hfs}(\omega)) - \omega\,D_T^j(\omega)\big] 
G(\omega + E_0)\,\big[ p^j(V_\mathrm{hfs}(\omega)) + \omega\,D_T^j(\omega)\big] | \phi \rbr\,.
\end{align}
Indeed, the recoil corrections take the compact form in the temporal gauge. It would be worthwhile
to derive them directly in this gauge, because a derivation of radiative recoil corrections
would otherwise be much  more complicated.

\subsection{Spin-orbit contribution}
The spin-orbit contribution 
\begin{align}
E_\mathrm{so} =&\ -\frac{q}{4\,M^2}\,(g-1)\,\vec I\cdot \langle\Psi| \vec E\times\vec\Pi-\vec\Pi\times\vec E|\Psi\rangle_\mathrm{QED} 
\nonumber \\ =&\ E_\mathrm{so1} + E_\mathrm{so2} + E_\mathrm{so3}\,, \label{52}
\end{align}
is split into three parts. In the $E_\mathrm{so3}$ part, $\vec \Pi \rightarrow -i\vec\partial_R$ and
\begin{align}
E_\mathrm{so3} =&\ i\,\frac{q}{4\,M^2}\,(g-1)\,\vec I\cdot\langle\Psi|[\vec E\times\vec\partial_R + \cev\partial_R\times\vec E |\Psi\rangle_\mathrm{QED}
\nonumber \\ =&\  
i\, \frac{q}{4\,M^2}\,(g-1)\,\epsilon^{ijk} I^i  \int d^4x\, \langle\Psi| \mathrm{T} [E^j(\vec R) \partial_R^k\,(-i\,e)\,j_\mu(x)\,A^\mu(x)] |\Psi\rangle + \mathrm{h.c.}
\nonumber \\ =&\ 
\frac{q\,e}{2\,M^2}\,(g-1)\,I^i \int \frac{d\,\omega}{2\,\pi\,i}\,\langle\phi | \epsilon^{ijk}\,\alpha^j\,\partial^k D(\omega) | \phi\rangle\,.
\end{align}
In the $E_\mathrm{so2}$ part, $\vec \Pi \rightarrow i \int d^3r\, \hat \psi^+(\vec r)\,\vec \partial_r\,\hat \psi(\vec r) $ and
\begin{align}
E_\mathrm{so2} =&\ -i\,\frac{q\,(g-1)}{2\,M^2}\, \epsilon^{ijk}\, I^i\,  \int d^3r\, \langle\Psi | E^j(\vec R) \, \hat \psi^+(\vec r)\,\partial^k_r\,\hat \psi(\vec r) |\Psi\rangle_\mathrm{QED}
\nonumber \\ =&\ 
-i\,\frac{q\,(g-1)}{2\,M^2}\, \epsilon^{ijk}\, I^i \int d^4x \int d^3r\, 
\langle\Psi | \mathrm{T}[E^j(\vec R) \, \hat \psi^+(\vec r)\,\partial^k_r\,\hat \psi(\vec r)\,(-i\,e)\,j_\mu(x)\,A^\mu(x)] |\Psi\rangle
\nonumber \\ =&\
-i\,\frac{q\,e\,(g-1)}{2\,M^2}\, \epsilon^{ijk}\, I^i \int \frac{d\omega}{2\,\pi}\,
\big[ \langle \phi | \partial^k\, G(E_0+\omega)\, \big[-\omega\,\alpha^l \,G_C^{lj}(\omega) +i\,\partial^jG_C^{00}\big]| \phi\rangle
\nonumber \\ &\
+ \langle \phi | \big[\omega\,\alpha^l \, G_C^{lj}(\omega) +i\,\partial^jG_C^{00}\big] \, G(E_0+\omega)\,\partial^k | \phi \rangle  \big]\,.
\end{align} 
In the $E_\mathrm{so1}$ part, $\vec \Pi \rightarrow -q\,\vec A(\vec R)$ and
\begin{align}
E_\mathrm{so1} =&\ \frac{q^2\,(g-1)}{2\,M^2}\,\vec I\cdot\,\langle\Psi |  \vec E(\vec R)\times\vec A(\vec R)  |\Psi\rangle_\mathrm{QED}
\nonumber \\ =&\
\frac{q^2\,(g-1)}{2\,M^2}\,\epsilon^{ijk}\,I^i\,\frac{(-i e)^2}{2}\, \int d^4x \int d^4y 
\langle\Psi |  \mathrm{T}[ E^j(\vec R) \, A^k(\vec R)\,j_\mu(x)\,A^\mu(x)\,j_\nu(y)\,A^\nu(y)] |\Psi\rangle
\nonumber \\ =&\
-\frac{e^2\,q^2\,(g-1)}{2\,M^2}\,\epsilon^{ijk}\,I^i\, \int \frac{d\omega}{2\,\pi}\,
\big[ \langle \phi | \big[ \omega\,\alpha^m\, G_C^{mj}(\omega) + i\,\partial^j G_C^{00}\big]\,G(E_0+\omega)\,\alpha^l\,G_C^{lk}(\omega) | \phi \rangle
\nonumber \\ &\
+ \langle \phi | \alpha^l\,G_C^{lk}(\omega)\,  G(E_0+\omega)\,\big[-\omega\,\alpha^m\, G_C^{mj}(\omega) + i\,\partial^j G_C^{00}\big] | \phi \rangle \big]\,.
\end{align}
The total spin-orbit part using Eq. (\ref{52}) is 
\begin{align}
E_\mathrm{so} =&\
-\frac{4\,\pi\,Z\,\alpha\,(g-1)}{2\,M^2}\,\epsilon^{ijk}\,I^i \int \frac{d\,\omega}{2\,\pi}\,\Bigl\{
-i\,\langle\phi | \alpha^j\,\partial^k D(\omega) | \phi\rangle
\nonumber \\ &\
+ \langle \phi | \big[ \omega\,\alpha^m\, G_C^{mj}(\omega) + i\,\partial^j G_C^{00}\big]\,G(E_0+\omega)\,\big[ p^k + 4\,\pi\,Z\,\alpha\,\alpha^l\,G_C^{lk}(\omega)\big] | \phi \rangle
\nonumber \\ &\
+ \langle \phi | \big[p^k + 4\,\pi\,Z\,\alpha\,\alpha^l\,G_C^{lk}(\omega)\big]\,  G(E_0+\omega)\,\big[-\omega\,\alpha^m\, G_C^{mj}(\omega) + i\,\partial^j G_C^{00}\big] | \phi \rangle
\Bigr\}\,.
\end{align}
It becomes simplified in the ${\cal F}_2$ form,
\begin{align}
E_\mathrm{so} =&\
\frac{(g-1)}{M^2}\,\epsilon^{ijk}\,I^i \int \frac{d\,\omega}{2\,\pi}\,\frac{1}{\omega}\,
\langle \phi | \big[ p^j(V_C) -\omega\,D_C^j(\omega) \big]\,G(E_0+\omega)\,\big[ p^k(V_C) +\omega\,D_C^k(\omega)\big] | \phi \rangle
\end{align}
and even more simplified in the ${\cal F}_3$ form,
\begin{align}
E_\mathrm{so} =&\
-\frac{(g-1)}{M^2}\,\epsilon^{ijk}\,I^i \int_s \frac{d\,\omega}{2\,\pi}\,\omega\,\langle \phi | D_T^j(\omega)\, G(E_0+\omega)\,D_T^k(\omega) | \phi \rangle\,.
\end{align}

\subsection{Double hfs contribution}
The double hfs contribution is due to the two-photon exchange between the bound electron and the nucleus magnetic moment $\vec \mu$ \cite{hfsdirac},
\begin{align}
E_{\rm sec} =&\
i\,e^2\,\int\frac{d\,\omega}{2\,\pi}\,\int\frac{d^3k_1}{(2\,\pi)^3}\,\int\frac{d^3k_2}{(2\,\pi)^3}\,
\frac{\rho(k_1^2-\omega^2)}{\omega^2-k_1^2+i\,\epsilon}\,
\frac{\rho(k_2^2-\omega^2)}{\omega^2-k_2^2+i\,\epsilon}\,
\langle\phi|\alpha^i\,e^{i\,\vec k_1\vec r}\,G(E_D+\omega)\,\alpha^j\,e^{-i\,\vec k_2\vec r}\, | \phi \rangle
\nonumber \\ & \times
\Bigl[ (\vec\mu\times\vec k_1)^{\,i}\,\frac{1}{-\omega+i\,\epsilon}\,(\vec\mu\times\vec k_2)^{\,j}
+ (\vec\mu\times\vec k_2)^{\,j}\,\frac{1}{\omega+i\,\epsilon}\,(\vec\mu\times\vec k_1)^{\,i}\Bigr]\,. 
\end{align}
We shall make at this point a note regarding the reference state singularity.
The reducible contribution in the ladder diagram, where the intermediate state is the same as the external one, $i\,\epsilon$ in the denominator of the 
fermion propagator should reverse the sign, namely $i\,\epsilon \rightarrow -i\,\epsilon$ (see Ref. \cite{shabaevPR}), which effectively leads
to a symmetric integration in $\omega$ around a pole at $\omega=0$.

Let us now decompose the product of the nuclear magnetic moments into irreducible parts, namely
\begin{align}
\mu^a\,\mu^b = \frac{\delta^{ab}}{3}\,\vec\mu^{\,2} + 
\frac{1}{2}\,\biggl(\mu^a\,\mu^b + \mu^b\,\mu^a-\frac{2\,\delta^{ab}}{3}\,\vec\mu^{\,2}\biggr) +
\frac{1}{2}\,\big[\mu^a\,,\,\mu^b\big]\,. \label{62}
\end{align}
Only the last part contributes to the magnetic dipole hyperfine splitting, and
\begin{align}
E_\mathrm{sec} =&\ -i\,e^2\,[\mu^i\,,\,\mu^j] \int_s \frac{d\,\omega}{2\,\pi}\,\frac{1}{\omega}\, \langle \phi |(\vec\alpha\times\vec \nabla)^i\,D(\omega,\vec r)\,
G(E_D+\omega)\,(\vec\alpha\times\vec \nabla)^j\,D(\omega,\vec r)\,| \phi \rangle \,.
\end{align}

\subsection{Final formula for recoil correction to hfs}
The total recoil correction to the hyperfine splitting in hydrogen-like ions is
\begin{align}
E_\mathrm{hfsrec} =&\ E_\mathrm{kin} + E_\mathrm{so} + E_\mathrm{sec}\,, \\
E_\mathrm{kin} = &\  
\frac{1}{M} \int_s \frac{d\omega}{2\,\pi}\,\frac{1}{\omega}\,
\bigl[ \lbr \phi | D_T^j(\omega) \,G(E_D+\omega)\, \partial^j(V_\mathrm{hfs}(\omega)) | \phi \rbr  
-  \lbr \phi | \partial^j(V_\mathrm{hfs}(\omega)) \,G(E_D+\omega)\, D_T^j(\omega) | \phi \rbr \bigr]
\nonumber \\ &\
+ \delta_\mathrm{hfs}\frac{i}{M} \int_s \frac{d\omega}{2\,\pi}\, \lbr \phi | D_T^j(\omega)\,G(E_D+\omega )\, D_T^j(\omega) | \phi \rbr\,, \label{65}
\\ 
E_\mathrm{so} = &\  
-\frac{(g-1)}{M^2}\,\epsilon^{ijk}\,I^i \int_s \frac{d\,\omega}{2\,\pi}\,\omega\,\langle \phi | D_T^j(\omega)\, G(E_D+\omega)\,D_T^k(\omega) | \phi \rangle\,, \label{66} \\
E_\mathrm{sec} = &\  
\biggl(\frac{4\,\pi\,Z\,\alpha}{2\,M}\,g\biggr)^2\,  \epsilon^{ijk}\,I^k  \int_s \frac{d\,\omega}{2\,\pi}\,\frac{1}{\omega}\, \langle \phi |(\vec\alpha\times\vec \nabla)^i\,D(\omega)\,
G(E_D+\omega)\,(\vec\alpha\times\vec \nabla)^j\,D(\omega)\,| \phi \rangle\,, \label{67}
\end{align}
where  $\vec D_T(\omega)$ is defined in  Eq. (\ref{13}), $V_\mathrm{hfs}(\omega)$ in Eq. (\ref{46}), and $D(\omega)$ in Eq. (\ref{47}). 
We can now replace $\rho$ by $\rho_C$ or $\rho_M$ depending on the presence of the $g$ factor. Namely in Eq. (\ref{65}) $\rho$ in
$D_T^j$ is replaced by $\rho_C$ and in $V_\mathrm{hfs}$ by $\rho_M$. 
In Eq. (\ref{66}) $\rho$ in $g\,D_T$ is replaced by $\rho_M$, while in the other $D_T$ is replaced by $\rho_C$. 
In Eq. (\ref{67}) $\rho$ in both $D$ are replaced by $\rho_M$.
In the next section we will verify this formula for $E_\mathrm{hfsrec}$  by the derivation of the $(Z\,\alpha)^5$ correction, which has already been obtained by other means in Ref. \cite{hfsdirac}.

\section{$(Z\,\alpha)^5$ recoil correction}
Let us at first calculate the  $(Z\,\alpha)^5$ finite nuclear size correction to the  energy given by Eq. (\ref{14}).
This correction comes from the hard two-photon exchange  and is split into two parts,
\begin{align}
E^{(5)}_\mathrm{rec} =&\ E^{(5)}_\mathrm{rec1} + E^{(5)}_\mathrm{rec2}\,.
\end{align}
In the first part these hard two exchanged photons are  $D_T(\omega)$, thus
\begin{align}
E^{(5)}_{\mathrm rec1} =&\ 
\frac{i}{M}\,\phi^2(0)\,(4\,\pi\,Z\,\alpha)^2\, \int_s \frac{d^4k}{(2\,\pi)^4}\,\frac{\bigl[\rho^2(-k^2)-1\bigr]}{(k^2)^2}\,
\Bigl(\delta^{ik} -\frac{k^i\,k^k}{\omega^2}\Bigr)\,\Bigl(\delta^{jk} -\frac{k^j\,k^k}{\omega^2}\Bigr)\,
\mathrm{Tr}\biggl[\gamma^i \,\frac{1}{(\not\!t+\not\!k-m)}\,\gamma^j\,\frac{(I+\gamma^0)}{4}\biggr]
\nonumber \\ =&\ 
\frac{1}{M}\,\phi^2(0)\,(4\,\pi\,Z\,\alpha)^2\, \int_s \frac{d^4k}{(2\,\pi)^4\,i}\,\frac{\bigl[\rho^2(-k^2)-1\bigr]}{k^4}\,
\biggl[\frac{2\,m\,(k^4+2\,\omega^4)}{\omega^2\,(k^4-4\,m^2\,\omega^2)}\biggr]
\nonumber \\ \stackrel{E}{=}&\ 
\frac{1}{M}\,\phi^2(0)\,(4\,\pi\,Z\,\alpha)^2\, \int_s \frac{d^4q}{(2\,\pi)^4}\,\frac{\bigl[\rho(q^2)-1\bigr]}{q^4}\,
A\biggl[  -\frac{2\,m\,(q^4+2\,q_0^4)}{q_0^2\,(q^4+4\,m^2\,q_0^2)}\biggr]\,, \label{69}
\end{align}
\end{widetext}
where in the last line we performed the Wick rotation, and
$A$ denotes an average over the three-dimensional sphere in the Euclidean space,
\begin{align}
A[f] \equiv \int \frac{d\,\Omega_q}{2\,\pi^2}\,f(q,q_0) = \frac{2}{\pi}\int_0^\pi d\phi\,\sin^2(\phi)\, f\big(q,q\,\cos(\phi)\big)\,,
\end{align}
then
\begin{align}
A\biggl[\frac{1}{q^4+4\,m^2\,q_0^2}\biggr] =&\  \frac{2}{q^4}\,\frac{1}{1 + \sqrt{1 + a^2}}\,,\\
A\biggl[\frac{1}{q_0^2}\biggr] =&\ -\frac{2}{q^2}\,,
\end{align}
where $a=2\,m/q$. In the second formula we assumed a symmetric integration around the pole at $q^0=0$, as denoted by subscript $s$ in Eq. (\ref{69}).
Applying this angle average
\begin{align}
E^{(5)}_\mathrm{rec1} =&\ 
\frac{m}{M}\,\phi^2(0)\,(Z\,\alpha)^2\,8\!\!\int_\mathrm{sub}\! \frac{dq}{q^3} \big[b - 1 - b^{-2} \big] \big[\rho^2(q^2)-1\big], \label{73}
\end{align}
where $b=1+\sqrt{1+a^2}$, and ``sub" denotes the subtraction of low $q$ singularity which corresponds to $(Z\,\alpha)^4$ finite size correction.
Equation (\ref{73}) agrees with the one derived previously in Ref. \cite{fsrec}'s Eq. (12) .  For electrons, it can be further simplified in terms of 
the effective radius $\langle r^2\,\ln(m\,r)\rangle$. 

The second part $E^{(5)}_\mathrm{rec2}$ comes from the nonrecoil hard two-Coulomb photon exchange 
\begin{align}
E^{(5)}_\mathrm{nrec} =&\ -\frac{\pi}{3}\,\phi^2(0)\,(Z\,\alpha)^2\,m\,r_F^3 \equiv \langle\phi| V^{(5)}_\mathrm{nrec} |\phi\rangle\,,
\end{align}
where
\begin{align}
r^3_F = \int d^3r_1\int d^3r_2\, \rho_C(r_1)\,\rho_C(r_2)\,|\vec r_1-\vec r_2|^3, 
\end{align}
and from the operator in Eq. (\ref{14}) replaced by the nonrelativistic nuclear kinetic energy [cf. Eq. (\ref{27})], so
\begin{align}
E^{(5)}_\mathrm{rec2} =&\ 2\,\langle\phi| V^{(5)}_\mathrm{nrec}\,\frac{1}{(E-H)'}\, \frac{\vec p\,^2}{2\,M} |\phi\rangle = -3\,\frac{m}{M}\, E^{(5)}_\mathrm{nrec}\,.
\end{align}
It can be interpreted as a reduced mass scaling of the nonrecoil $(Z\,\alpha)^5$ correction.

We are now ready to pass to  the $(Z\,\alpha)^5$ hyperfine recoil correction. It also split into two parts,
\begin{align}
E^{(5)}_\mathrm{hfsrec} =&\ E^{(5)}_\mathrm{hfsrec1} + E^{(5)}_\mathrm{hfsrec2}\,.
\end{align}
The first part comes from the hard two-photon exchange, where these two photons are $D_T(\omega)$ and $D(\omega)$ or $V_\mathrm{hfs}(\omega)$,
but here we do not subtract the point nucleus contribution. Following closely the previous case of $E^{(5)}_\mathrm{rec1}$ we obtain
\begin{align}
E^{(5)}_\mathrm{hfsrec1} =&\ 
-\frac{16}{3}\,\frac{(Z\,\alpha)^2}{M^2}\,\phi^2(0)\,\vec I\cdot\vec s \int_\mathrm{sub} \frac{d q}{q}
\biggl[\rho_C^2\,\biggl(\frac{2}{b}+\frac{1}{2\,b^2}\biggr)
\nonumber \\ &\hspace*{-8ex}
+\frac{g}{2}\,\rho_M\,\rho_C\,\biggl(2\,(b-1) - \frac{2}{b} -\frac{1}{b^2}\biggr)
+\frac{g^2}{4}\,\rho_M^2\,\biggl( \frac{1}{2\,b^2} - \frac{1}{b} \biggr)\biggr].
\end{align}
It requires low $q$ subtraction as denoted by ``sub", which corresponds to the leading hyperfine splitting of the order of $(Z\,\alpha)^4$. 
Namely, the term $2\,(b-1)$ contains the linear singularity $2\,(b-1)\sim 2\,a = 4\,m/q$ for the small $q$, which should be subtracted out.
Then, $2\,(b-1-a) = 2/(a+\sqrt{1+a^2}) \equiv 2/b'$ and
\begin{align}
E^{(5)}_\mathrm{hfsrec1} =&\ 
-\frac{16}{3}\,\frac{(Z\,\alpha)^2}{M^2}\,\phi^2(0)\,\vec I\cdot\vec s \int_0^\infty \frac{d q}{q}\,
\biggl[\rho_C^2\,\biggl(\frac{2}{b}+\frac{1}{2\,b^2}\biggr)
\nonumber \\ &\hspace*{-2ex}
+\frac{g}{2}\,\rho_M\,\rho_C\,\biggl(\frac{2}{b'} - \frac{2}{b} -\frac{1}{b^2}\biggr)
+\frac{g^2}{4}\,\rho_M^2\,\biggl( \frac{1}{2\,b^2} - \frac{1}{b} \biggr)\biggr] 
\nonumber \\ &\hspace*{-2ex}+ \delta E^{(5)}_\mathrm{hfsrec1}
\end{align}
in agreement with Ref. \cite{hfsdirac}. The last term $\delta E^{(5)}_\mathrm{hfsrec1}$  is
\begin{align}
\delta E^{(5)}_\mathrm{hfsrec1} =&\ 
\frac{16\,\pi}{3}\,\frac{(Z\,\alpha)^2}{M^2}\,\phi^2(0)\,\vec I\cdot\vec s\; \frac{g}{2}\,m\,r_Z \,,
\end{align}
where
\begin{align}
r_Z =&\ \frac{1}{\pi^2}\,\int\frac{d^3q}{q^4}\,\big[1-\rho_C(q^2)\,\rho_M(q^2)\big]
\end{align}
is the so-called Zemach radius \cite{Zemach}.

The second part $E^{(5)}_\mathrm{hfsrec2}$ comes from the second term in Eq (\ref{65}),  
where the perturbation is due to the nonrecoil hyperfine correction
\begin{align}
E^{(5)}_\mathrm{hfsnrec} =&\ -\frac{16\,\pi}{3}\,(Z\,\alpha)^2\,\phi^2(0)\,\frac{g}{2\,M}\,\vec I\cdot\vec s\,r_Z 
\nonumber \\ \equiv&\
 \langle\phi| V_\mathrm{hfsnrec} |\phi\rangle
\end{align}
and with recoil replaced by the nuclear kinetic energy
\begin{align}
E^{(5)}_\mathrm{hfsrec2}  =&\  2\,\langle\phi| V_\mathrm{hfsnrec}\,\frac{1}{(E-H)'} \frac{\vec p\,^2}{2\,M}|\phi\rangle 
\nonumber \\ =&\ -3\,\frac{m}{M}\,E^{(5)}_\mathrm{hfsnrec}\,.
\end{align}
It can be interpreted as a reduced mass scaling of $E^{(5)}_\mathrm{hfsnrec}$. Together with 
\begin{align}
\delta E^{(5)}_\mathrm{hfsrec1} = -m/M\,E^{(5)}_\mathrm{hfsnrec}
\end{align}
it gives the factor $4$ and the total reduced mass scaling $(\mu/m)^4$ of the Zemach contribution $E^{(5)}_\mathrm{hfsnrec}$, 
in agreement with Ref. \cite{hfsdirac}.

\section{Summary}
We have introduced a general quantum electrodynamic method for the derivation of nuclear recoil corrections in hydrogenic systems,
and we present an exemplary derivation of the nuclear recoil correction to the hyperfine splitting. 
The exact formulas in $Z\,\alpha$ are shown in Eqs. (\ref{65}) - (\ref{67}). They can be used for the direct numerical calculation of the nuclear
recoil effects, or for an analytic derivation of $Z\,\alpha$ expansion coefficients, in particular of the $O(Z\,\alpha)^2\,E_F$ contribution, which was
originally derived by Bodwin and Yennie in Ref. \cite{Bodwin:88}, but has not been confirmed.

This general method can be applied for the derivation of all the other nuclear recoil effects of an arbitrary order in the mass ratio,
including radiative recoil. It would be worthwhile, however, to simplify the derivation by direct use of the temporal gauge, because
formulas are very much simplified in this gauge.
 
\appendix
 \section{Photon propagator}
The photon propagator in the Feynman gauge is
\begin{align}
i\,G^{\mu\nu}(x'-x) =&\ \langle0|{\mathrm T} A^\mu(x')\,A^\nu(x)|0\rangle
\nonumber \\ =&\
-i\,g^{\mu\nu}\int\frac{d^4 k}{(2\,\pi)^4}\frac{e^{-i\,k(x'-x)}}{k^2+i\,\epsilon} \,,
\\ G^{\mu\nu}(k)=&\ -\frac{g^{\mu\nu}}{k^2+i\,\epsilon}\,,
\end{align}
while in a Coulomb gauge with finite size,
\begin{align}
G_C^{00} =&\ \rho(\vec k^2)/\vec k^2, \\
G_C^{ij}(k) =&\ \frac{\rho(-k^2)}{k^2}\,\biggl(\delta^{ij}-\frac{k^i\,k^j}{(k^0)^2}\biggr) - \frac{k^i\,k^j}{(k^0)^2}\,\frac{\rho(\vec k^2)}{\vec k^2}\,.
\end{align}
The transverse part is not orthogonal to $k^i$
\begin{align}
k^i\,G_C^{ij}(k) =&\ \bigl[ \rho(-k^2) - \rho(\vec k^2)\bigr]\,\frac{k^j}{(k^0)^2}\,,
\end{align}
in contrast to the propagator in the regular Coulomb gauge.
The photon propagator in the temporal gauge with finite size is
\begin{align}
G_T^{ij}(\omega, \vec k) =&\ \frac{\rho(-k^2)}{k^2}\,\biggl(\delta^{ij}-\frac{k^i\,k^j}{\omega^2}\biggr)\,.
\end{align}

The auxiliary propagators are
\begin{align}
\langle0|{\mathrm T} A^i(x)\,B^j(y)|0\rangle =&\ -i\,\epsilon^{ijk}\,\nabla^k\,{D}(x-y)\,,\\
D(x-y) =&\ \int \frac{d^4k}{(2\pi)^4}\, e^{-i\,k\,r}\,\frac{\rho(-k^2)}{k^2}\,,\\
D(\omega,r) =&\ \int \frac{d^3k}{(2\pi)^3}\, e^{i\vec{k}\cdot\vec{r}}\,\frac{\rho({\vec k}^2-\omega^2)}{\omega^2-{\vec k}^2}\,, \\
V_\mathrm{hfs}(\omega,r) =&\  \epsilon^{ijl}\, e\,\mu^i\, \alpha^j\, \partial^l\, D(\omega, r)\\ 
=&\ -\frac{4\,\pi\,Z\,\alpha}{2\,M}\,g\, \vec I\cdot \vec \alpha\times \vec\nabla D(\omega, r). \nonumber 
\end{align}

\section{Spinor field}
The representation of the spinor field in terms of solutions of the Dirac equation are
\begin{align}
\hat\psi(x)=&\ \sum_s^+ a_s\phi_s(\vec x)\,e^{-i\,E_s t} + \sum_s^- b_s\phi_s(\vec x)\,e^{-i\,E_s t}\,,
\nonumber \\
\hat\psi^+(x)=&\ \sum_s^+ a^+_s\phi^+_s(\vec x)\,e^{i\,E_s t} + \sum_s^- b^+_s\phi^+_s(\vec x)\,e^{i\,E_s t}\,,
\end{align}
which form a complete basis
\begin{align}
\sum_s^+ \phi_s(x')\,\phi^+_s(x) + \sum_s^- \phi_s(x')\,\phi^+_s(x) =&\ \delta^3(\vec x' -\vec x)\,.
\end{align}
The fermionic anticommutation relations are
\begin{align}
\{a_r\,,\,a^+_s\} = \{b_r\,,\,b^+_s\} = \delta_{r,s}\,,
\end{align}
and
\begin{align}
\{\hat\psi(\vec x\,',0)\,,\,\hat\psi^+(\vec x,0)\} =&\  \delta^3(\vec x' -\vec x)\,.
\end{align}
The fefinition of the vacuum state is
\begin{align}
a_r|0\rangle = b^+_r|0\rangle = 0\,.
\end{align}
The projection operators into positive and negative energy subspace are
\begin{align}
P^+ =&\ \sum_s^+ \phi_s(x')\,\phi^+_s(x) = \langle0|\hat\psi(\vec x\,',0)\,\hat\psi^+(\vec x,0)|0\rangle\,,
\nonumber \\
P^- =&\ \sum_s^- \phi_s(x')\,\phi^+_s(x) = \langle0|\hat\psi^+(\vec x,0)\,\hat\psi(\vec x\,',0)|0\rangle\,,
\end{align}
with
\begin{align}
P^++P^- =&\ I\,.
\end{align}
The fermion propagator is
\begin{align}
i\,G(x',x) =&\ \langle0|{\mathrm T}\hat\psi(x')\,\hat\psi^+(x)|0\rangle
\nonumber \\ =&\ 
i\,\int\frac{d\omega}{2\,\pi}\sum_s\frac{\psi_s(\vec x\,')\,\psi^+_s(\vec x)}{\omega - E_s(1-i\,\epsilon)}\,e^{-i\,\omega\,(x'^0-x^0)} \,.
\end{align}
The equal time propagator can be written as
\begin{align}
i\,G(x',x)|_{x'^0=x^0} =&\ \langle0|{\mathrm T}\hat\psi(x')\,\hat\psi^+(x)|0\rangle|_{x'^0=x^0}
\nonumber \\ =&\ 
\frac{1}{2}\,\langle0|\hat\psi(x')\,\hat\psi^+(x) - \hat\psi^+(x)\,\hat\psi(x') |0\rangle
\nonumber \\ =&\ \frac{1}{2}\,(P^+-P^-)\,.
\end{align}

\end{document}